\documentclass[12pt]{article}

\usepackage{graphicx,epsfig}
\usepackage{color}
\usepackage{multirow}

\def\be{\begin{equation}}
\def\ee{\end{equation}}
\def\ba{\begin{eqnarray}}
\def\ea{\end{eqnarray}}
\def\L{{\cal L}}

\definecolor{verde}{rgb}{0.4,0.6,0}
\definecolor{azul}{rgb}{0.1,0.2,0.6}
\definecolor{rojo}{rgb}{0.8,0.1,0.1}

\def\verde{\color{verde}}
\def\rojo{\color{rojo}}
\def\azul{\color{azul}}

\begin{document}

\begin{center}

{\Large{\bf Constraining Lorentz violations with Gamma Ray Bursts} }

\bigskip \bigskip

{\large Mar\'{\i}a Rodr\'{\i}guez Mart\'{\i}nez and Tsvi Piran}

\bigskip

Racah Institute of Physics  \\
The Hebrew University \\
91904 Jerusalem, Israel.

 \end{center}

\bigskip

\begin{abstract}

  Gamma ray bursts are excellent candidates to constrain physical  models which
  break the Lorentz symmetry. We consider deformed dispersion
  relations which break boost invariance and lead to an
  energy-dependent speed of light. In these models, simultaneously
  emitted photons from cosmological sources reach Earth with a
  spectral time delay that depends on the symmetry breaking scale. We
  estimate the possible bounds which can be obtained by comparing the
  spectral time delays with the time resolution of  available
  telescopes. We discuss the best strategy to reach the strongest
  bounds. We compute the probability of detecting bursts that improve
  the current bounds. The results are encouraging, depending on the
  model, it is possible to build a detector that within several years
  will improve the present limits of 0.015 $m_{pl}$.

\end{abstract}


\setcounter{equation}{0}
\section{Introduction}

One of the open questions of high energy physics is how to unify
gravity with quantum physics. A lot of effort has been devoted to
develop a theory of quantum gravity. This theory is likely to
require a drastic modification of our current understanding of the
space-time.  At present there are two formal mathematical
approaches : loop quantum gravity and superstring theory. Whatever
might be the right description of the space-time at very short
scales, there are some likely physical manifestations. It has been
suggested, for instance, that such a theory would break what we
believe to be basic symmetries of nature. In
\cite{Wheeler:1957mu,Hawking:1978}, it was shown that Einstein
Lagrangian allows for large fluctuations of the metric and the
topology of the space-time on scales of order of the Planck
length, creating a foam-like structure at these scales. It has
been proposed that the propagation of particles in a foamy
space-time is strongly affected on short scales. The medium
responds differently depending on the energy of the particle, in
analogy with the propagation through a conventional
electromagnetic plasma \cite{Drummond:1979pp,Latorre:1994}. Thus
space-time might exhibit a non-trivial dispersion relation in
vacuum, violating therefore Lorentz invariance.

There are many different ways of breaking Lorentz symmetry; a
background tensor field, like a magnetic field, breaks the vacuum
rotational invariance, for instance. However, it has been shown that
there are 46 different ways in which the standard model Lagrangian can
be modified while remaining renormalizable, invariant under $SU(3)
\times SU(2) \times U(1)$ and rotationally and translationally
invariant in a preferred frame \cite{Coleman:1998ti}. Among other
effects, these terms cause the velocity of light to differ from the
maximum attainable velocity of a particle, therefore changing the
kinematics of particle decays. Modifying the dispersion relations of
photons and electrons allows for new QED vertex interactions like
photon splitting in vacuum, vacuum \v{C}erenkov effect for electrons,
photon decay, electron-positron annihilation to a single photon, etc.
\cite{Jacobson:2002hd,mewes}.

In this work we consider only rotationally invariant deformations of
the photon dispersion relation which produce an energy-dependent speed
of light. If this effect is present in nature, it has to be absent
below some energy scale, $\xi \, m_{pl}$, (where $m_{pl}$ is the
Planck energy and $\xi$ a dimensionless constant), high enough to have
been undetected so far.  Traditionally this energy is believed to be
the Planck energy which, at first, seems to make hopeless any
experimental attempt to test these models. However, in 1997
Amelino-Camelia et al. \cite{Amelino-Camelia:1997gz} suggested that
such models can be explored by studying the propagation of photons
emitted from a distance source like a gamma ray burst (GRB) GRBs are
short and intense pulses of soft gamma rays that arrive from
cosmological distances from random directions in the sky. The bursts
last from a fraction of a second to several hundred seconds. Most GRBs
are narrowly beamed with typical energies around $10^{51}$ ergs, making
them comparable to supernovae (for a recent review see
\cite{Piran:2004ba,Zhang:2003uk}). Because of the large distances
traveled by the photons, these bursts are valuable tools to explore
energies far beyond the reach of any laboratory on Earth.

If one considers photons with energies much smaller than the
symmetry breaking scale, it is possible to expand the dispersion
relation in powers of $E /\xi m_{pl}$. The first correction
produces a tiny departure from the Lorentz invariant (LI)
equations and we expect that the low energy limit of the deformed
dispersion relation can be written generically as:
\be E^2 - p^2  c^2 \simeq  \epsilon \, E^2 \left( {E \over \xi
m_{pl} }\right)^n , \label{mod-dis-rel}
\ee
where $\epsilon = \pm 1$ takes into account the possibility of having
either infraluminal or superluminal motion, the latter appearing in
some models of quantum loop gravity
\cite{Gambini:1998it,Alfaro:1999wd}. Photons simultaneously emitted
from a GRB with different energies will travel at different speeds,
and therefore will show on Earth a time delay.  The goal of this paper
is to explore these high energy Lorentz violation models by studying
such time delays. We analyze the potential of detecting observational
consequences of a modified dispersion relation like Eq.
\ref{mod-dis-rel}.

The paper is organized as follows; in section \ref{stand-th} we review
cosmological photon propagation in the LI theory and show how these
results are modified when a non LI term is introduced. We compute the
travel time of cosmological photons and, as a check, compare it with
the travel time obtained in the Newtonian approximation. In section
\ref{det-photons} we turn to the observations and show how such models
can be tested using GRB observations. We compare our method with
previous works in section \ref{comparacion} and finally conclude in
section \ref{conclusion}.

\section{Propagation of the photons}

\label{stand-th}

We consider first the propagation of a particle in a FRW universe,
described by the metric $ ds^2 = -c^2 \, dt ^2 + a(t)^2 d
\vec{x}^2$. The Hamiltonian of a relativistic particle is
\be
{\cal H} = \sqrt{m^2 \, c^4 + {p^2 \, c^2\over a^2}} \; ,
\label{hamiltonian}
\ee
where $p$ is the comoving momentum and $m$ the mass. The Hamiltonian
depends explicitly on $t$ through $a(t)$, expressing the fact that the
momentum is redshifted due to the cosmological expansion.  The
trajectory of the particle is
\be
x(t,p) = \int {p \, c^2 \over a^2}
{d t \over \sqrt{m^2 \, c^4+ {p^2 \, c^2\over a^2}}} \; \; \; ,
\; \;\; p = \mbox{constant} \; .
\label{distance}
\ee
For a  massless particle, like  a photon, Eq. \ref{distance}
becomes
\be
a \, \dot x = c \;.
\ee
Hence the speed of photons is an universal constant, $c$, which does
not depend on the energy.

When Lorentz symmetry is broken this result is modified. As long
as a theory of quantum gravity is not available, the high energy
corrections to the Hamiltonian defined in Eq. \ref{hamiltonian}
cannot be calculated. Here we will adopt a phenomenological
approach, assuming that the Hamiltonian at high energy is an
unknown function of the momentum, which reduces at low energies to
Eq. \ref{hamiltonian}. At small energies compared to the symmetry
breaking scale, $E \ll \xi \, m_{pl}$, a series expansion is
applicable. We will consider the first order correction to the LI
theory.

We consider models which break boost invariance but keep
rotational and translational invariance.  Inspired by Eq.
\ref{hamiltonian} we therefore postulate
\be {\cal H}  = \left[ m^2 c^4+ {p^2  c^2 \over a^2} \left(1 +
\left( {p \, c \over  \xi m_{pl} \, a} \right)^n  \right)
\right]^{1/2}
 , \;\;\; n = 1, 2,... \;\;\;.
\label{mod_disp_rel}
\ee
Note that given Eq. \ref{mod-dis-rel} there is some arbitrariness in
the choice of Eq. \ref{mod_disp_rel} concerning the $a^{-n}$ factor.
We believe that this choice is physically the best motivated because
any $p$ dependence should be redshifted due to the cosmic expansion.

We define the parameter $\mu$ as the ratio between the energy of
the photon and the Planck energy,
\be \mu(a,p) = {p \, c \over  m_{pl} \, a} \; . \ee
From Eq. \ref{mod_disp_rel} we deduce the new trajectory of a particle
\ba
 x(t,p) & \simeq &
\int_{t_i}^{t} \frac{p \, c^2} {a^2 \sqrt{m^2  c^4 + {p^2  c^2\over a^2}}}
\left[ 1 + {1 \over 2} \, \mu^n \, \xi^{-n} \left(
  1 + n + {m^2 c^4 \over m^2 c^4  + { p^2 c^2 \over a^2 }} \right) \right ] dt
\; , \nonumber \\  p &=& \mbox{constant} \; .
\label{dist-mod} 
\ea
We made a linear expansion with respect to $\mu^n $ since Eq.
\ref{mod_disp_rel} is only valid to linear order.  The linear
approximation used in equation \ref{mod_disp_rel} remains valid
for $\mu < 1$. Notice that in the limit when the symmetry breaking scale goes
to infinity, $\xi \rightarrow \infty$, we recover Eq.
\ref{distance} and the Lorentz symmetry is restored.

The main and striking difference with respect to the LI theory is that
the speed of a massless particle {\it depends on its momentum}
\be
a \, \dot x = c (1 + {1\over 2} (1+n) \, \mu(a,p)^n \, \xi^{-n} )\;.
\label{energydependenceC}
\ee

\subsection{Time delay}

Because of the energy dependence of the speed of light in Eq.
\ref{energydependenceC}, two photons emitted at the same time from
the same source with momenta $p_1$ and $p_2$, will reach Earth at
times $t_1$ and $t_2$. The comoving distance traveled by both
photons is the same
\ba
x(t_1,p_1) &=&  x(t_2,p_2)  \label{eq-dist} \; , \\ \nonumber
t_2 & = &  t_1 + \Delta t \; .
\ea
Rewriting equation \ref{dist-mod} in terms of the redshift and
particularizing for photons, $m=0$, one obtains
\be
x(z,p) = {c \over H_0}\int_0^z  \left(1  + {1 +n \over 2} \,
\mu_0(p)^n \, \xi^{-n} \, (1+z)^n \right)
{ dz \over \sqrt{ \Omega_m  (1+z)^3 +  \Omega_\Lambda}} \; ,
\label{tray}
\ee
where the cosmological parameters $H_0$, $\Omega_m$ and
$\Omega_\Lambda$ are evaluated today \cite{Spergel:2003cb} and we set
$a_0 =1$. Notice that
\be
\mu (a,p) =  \mu_0 (p) \, (1+z)   \;\; , \;\; \mu_0 (p) \equiv \mu(a_0,p) \; .
\ee
When $\xi \rightarrow \infty$, Eq.  \ref{tray} becomes the
standard definition of the cosmological distance, as it should.

\begin{figure}
\begin{center}
\includegraphics{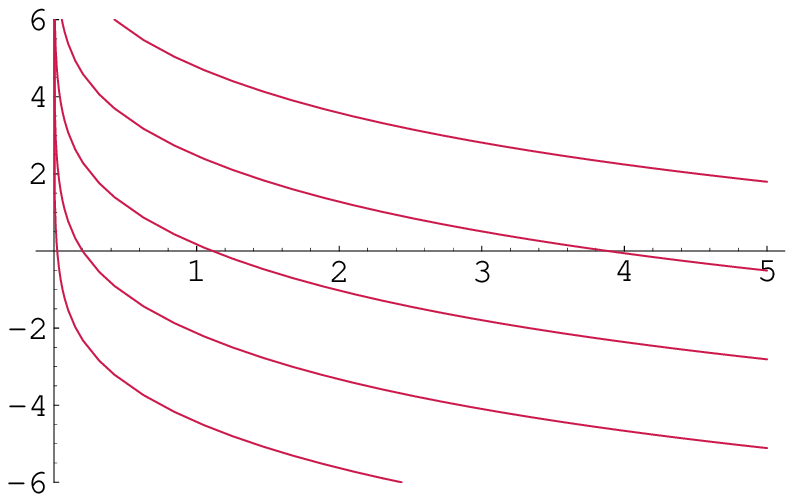}
\put(5,68){\azul $ z$ } \put(-233,155){\azul log E (GeV)}
\put(-109,2){\verde $10^{-3}$ s} \put(-2,13){\verde $10^{-2}$ s }
\put(-2,37){\verde 0.1 s} \put(-150,95){\verde 1 s }
\put(-123,115){\verde 10 s } \caption{Curves of constant time
delay as a tunction of the redshift of the source and the energy
of the photon. In all the curves $n =1$ and $\xi =1 $.} \label{cont-fig}
\end{center}
\end{figure}

Expanding Eq. \ref{eq-dist} for small $\Delta t$, we obtain the
time delay between two photons with momenta  $p_1$ and $p_2$,
\be 
\Delta t_{\mbox{{\scriptsize del}}} \simeq {1+n \over 2 H_0 \,
  \xi^n} \, \Delta \mu^n \, { \sqrt{ \Omega_m (1+z)^3 +
    \Omega_\Lambda}\over \sqrt{ \Omega_m + \Omega_\Lambda}} \int_0^z {
  (1+z)^n dz \over \sqrt{ \Omega_m (1+z)^3 + \Omega_\Lambda}} \; ,
\label{time-delay}
\ee
where $\Delta \mu^n = \mu_0^n (p_2) - \mu_0^n (p_1) $. Fig.
\ref{cont-fig} depicts the time delay as a function of the
redshift and the momentum of the photon.

\subsection{Newtonian approximation}

It is instructive to compare the former results with those obtained in
the Newtonian approximation. For small redshifts one can neglect the
expansion of the universe and suppose that energy of the photons is
constant. The delay between a low energy photon traveling at the
standard speed of light $c$ and the a high energy photon traveling at
the modified speed of light $v$ is
\be
\Delta t \sim {d \over c}- {d \over v} \;,
\ee
where $ v = {d E / dp}$.  In this approximation a linear relation
between distance and redshift holds
\be
d \simeq {c \over H_0 }
{z \over \sqrt{\Omega_m + \Omega_\Lambda}} \; .
\ee
Hence the time delay is
\be
\Delta t \sim  {1+n \over 2 H_0}\,{z \over \sqrt{\Omega_m + \Omega_\Lambda}}\,
 \mu_0^n \, \xi^{-n} \; .
\label{delay-aprox}
\ee
Comparing Eq. \ref{delay-aprox} and Eq. \ref{time-delay} we verify
that the Newtonian analysis agrees to first order with its
relativistic counterpart, Eq.  \ref{time-delay}.

\section{Observational Detection of Lorentz Violation}

\label{det-photons}

To obtain an experimental bound on the symmetry breaking scale
$\xi$, we need to compare the delay produced by the modified speed
of light, Eq.  \ref{energydependenceC}, with the time resolution
of the observing telescope. For a successful detection of a
Lorentz violation, the delay has to be larger than the time
resolution of the telescope:
\be 
\Delta t_{\mbox{{\scriptsize del}}} > \Delta t_{\mbox{{\scriptsize res}}}
\; . 
\label{bound} 
\ee

\subsection{Telescope time resolution}

The time resolution of a telescope depends on two factors. The
first is the intrinsic detector  minimal time resolution, $\Delta
t_{\mbox{{\scriptsize detector}}}$, which is typically of order
$10^{-3}-10^{-4}$s. The second factor, $\Delta
t_{\mbox{{\scriptsize res}}}$, is inversely proportional to the
photon detection rate which, in turn, depends on the detector
effective area and on the luminosity, spectrum and distance of the
source,
\be \Delta t_{\mbox{{\scriptsize res}}} \simeq   {b \over A \,
P_{\gamma}(E_1,E_2)} \; . \ee
where $P_\gamma(E_1,E_2)$ is the photon peak flux in the energy
band $(E_1,E_2)$ and $A$ is the detector effective area. The
factor $b$ takes into account the minimum number of photons needed
to resolve a peak. If the noise of the detector is negligible, $b$
is of order 5-10. In the following examples we consider an
idealized detector with no noise and set $b=6$.

The overall time resolution is given by
\be \tilde{\Delta t_{\mbox{{\scriptsize res}}}} = \max (\Delta
t_{\mbox{{\scriptsize res}}} ,\Delta t_{\mbox{{\scriptsize
detector}}} ) \; . \ee

The photon flux depends on the energy interval and therefore it is
sensitive to the spectrum of the burst. A good phenomenological
fit for the high energy GRB photon spectrum is
\be N ({\cal E}) = R_1 \, {\cal E}^{- \beta} \; , \label{spectrum}
\ee
where ${\cal E}$ is the energy in the rest frame of the
burst\footnote{The notation in this section is the following:
${\cal E}$ denotes energies in the rest frame of the burst and $E$
energies measured in the Earth frame.}, and $N({\cal E})$ has
units of photons keV$^{-1}$ s$^{-1}$.  This fit is valid for
energies higher than ${\cal E}_0$, where ${\cal E}_0$ is typically
of order of a few hundred keV. In what follows, we will take
${\cal E}_0 \sim $ 100 keV. For this spectrum the  flux in the
energy band $(E_1,E_2)$ is
\be P_\gamma(E_1,E_2) =  {1 \over 4 \pi d(z)^2} \, {R_1 \over 1+z}
\int^{E_2 (1+z)}_{E_1 (1+z)} {\cal E}^{-\beta} \,  d{\cal E} .
\label{part-detected} \ee
The factors $(1+z)$ in the limits of the integral transform rest-frame
energies into Earth measured energies, and there is an extra $(1+z)$
because of the cosmological time dilation; $d(z)$ is the cosmological
distance. 

Let us introduce the (isotropic equivalent) peak luminosity in the GRB
rest frame is
\be {\cal L}_{peak} =  R_1   \int^{\infty}_{{\cal E}_0} {\cal E}^{-\beta}
\, {\cal E}\,  d{\cal E} . 
\label{lumlum}
\ee
Combining Eq. \ref{part-detected} and Eq. \ref{lumlum} we obtain
\be \Delta t_{\mbox{{\scriptsize res}}} = b \, { 4 \pi d(z)^2
\over A} \, {\beta - 1 \over \beta -2} \, {(1+z)^{\beta } \over
{\cal L}_{peak} } \, {{\cal E}_0^{2-\beta} \over
E_1^{1-\beta}-E_2^{1-\beta} } \; , \label{time-res} \ee

In the following sections we will also need to estimate the
resolution from the photon energy flux $P(E_1,E_2)$ and from the
luminosity of the bursts. Using
\be P(E_1,E_2) =  {R_1 \over 4 \pi d(z)^2 \, (1+z)} \,  \int^{E_2
(1+z)}_{E_1 (1+z)} {\cal E}^{-\beta} \, {\cal E}\, d{\cal E} \;,
\label{energy-flux} \ee
the time resolution is obtained by combining Eq.
\ref{part-detected}, \ref{time-res} and \ref{energy-flux},
\be \Delta t_{\mbox{{\scriptsize res}}}\simeq {b \over A \,
P(E_1,E_2)}\, {\beta -1 \over \beta -2} \, (1+z) \,
{E_2^{2-\beta}-  E_1^{2-\beta} \over E_2^{1-\beta}-  E_1^{1-\beta}
} \; . \ee

\subsection{A bound on $\xi$}

\label{conditions-energy-redshift}

Comparing Eq. \ref{time-delay} and \ref{time-res} we find  that we
can test symmetry breaking scales up to:
\be \xi^n < {A H_0\over 8 \pi \, b\, c^2} \, {\beta - 2 \over
\beta -1} \, {\cal E}_0^{\beta-2} \, {\cal L}_{peak}  \, \left( 1
- \left(E_2 \over E_1\right)^{1-\beta} \right) \, {\Delta \mu^n
\over E_1^{ \beta-1}}  \, \, {\cal G}_n (z)  \; , \label{bound2}
\ee
where all the redshift dependence is contained in the function
${\cal G}_n (z)$ (remember that ${\cal L}_{peak}$ is defined in
the rest frame of the burst).

\begin{figure}
\begin{center}
\includegraphics{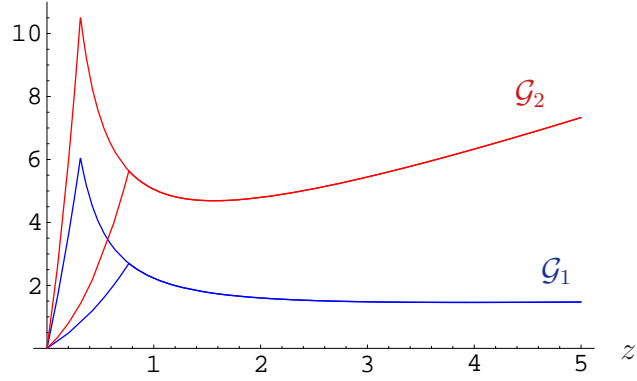}
\put(5,8){$ z$}
\put(-25,38){\azul ${\cal G}_1$}
\put(-35,105){\rojo ${\cal G}_2$}
\caption{${\cal G}_n(z)$ as a function of the redshift for $\beta =
  2.5$. For small redshifts ${\cal G}_n (z)$ has two branches, the
  upper one corresponds to $\Delta t_{\mbox{{\scriptsize detector}}} =
  10^{-4}$ s and the lower one corresponds to 
$\Delta t_{\mbox{{\scriptsize detector}}} =10^{-3}$ s.}
\label{funG}
\end{center}
\end{figure}

For distant bursts, the limiting resolution is determined by the
photon arrival rate and therefore $\tilde{\Delta
t_{\mbox{{\scriptsize res}}}} = \Delta t_{\mbox{{\scriptsize
res}}}$. In this case, the function ${\cal G}_n (z)$ is given by:
\ba
{\cal G}_n (z) &=& (1+n) { \sqrt{ \Omega_m  (1+z)^3 +  \Omega_\Lambda}\over
\sqrt{ \Omega_m  +  \Omega_\Lambda}} \,
\int_0^{z_0} { (1+z)^n dz \over \sqrt{ \Omega_m  (1+z)^3 +
    \Omega_\Lambda}}
\nonumber \\
&& \cdot \left( \int^z_0 {dz \over \sqrt{\Omega_m \, (1+z)^3 +
\Omega_\Lambda}}  \right)^{-2}  \, (1+z)^{-\beta } \; .
\label{redshiftdependence}
\ea
On the other hand, for nearby bursts $\tilde{\Delta
t_{\mbox{{\scriptsize res}}}}$ is limited by the detector
resolution. Fig. \ref{funG} depicts the behavior of ${\cal G}_n
(z)$. For small redshifts, ${\cal G}_n (z)$ increases from zero up to
a maximum value where $\Delta t_{\mbox{{\scriptsize detector}}}=
\Delta t_{\mbox{{\scriptsize res}}}$. ${\cal G}_n (z)$ is shown in
this regime for $\Delta t_{\mbox{{\scriptsize detector}}} = 10^{-4}$ s
(upper branch) and $10^{-3}$ s (lower branch).  For higher redshifts,
$\tilde{\Delta t_{\mbox{{\scriptsize res}}}}$ is dominated by $\Delta
t_{\mbox{{\scriptsize res}}}$. The function ${\cal G}_n (z)$ decreases
then up to a minimum and finally increases again at high redshift (for
$n=1$ the growth begins at $z>5$, which is too large to be
experimentally interesting). As the limit on $\xi$ is proportional to
${\cal G}_n^{1/n} (z)$ the best limit is obtained at small redshifts,
when $\Delta t_{\mbox{{\scriptsize res}}} = \Delta
t_{\mbox{{\scriptsize detector}}}$.

This is a non intuitive result. One would expect that the best
bound on $\xi$ is obtained from photons arriving from bursts at
very high redshifts, which have significant larger delays than
photons from nearby bursts. However the distance also dilutes the
photons decreasing the time resolution.  Moreover, due to the
redshift of the energy, a fixed energy band on Earth corresponds
to an intrinsically higher energy band which is more scarce in
photons.  These two combined effects overcome the improvement of
the bound due to the larger delay. Thus, to obtain the tightest
bound {\it it is preferable to use low redshift bursts}. As it can
be seen in Fig. \ref{funG}, low redshift means here $z<1$ (but
this depends, of course, on the intrinsic detector resolution).
Note that we have not yet taken into account the possible
attenuation of the interstellar medium. This result is merely
based on geometric considerations.

Let us turn our attention to the influence of the energy range in
which we observe the burst. If we observe photons in the interval
$(E_1,E_2)$, $\mu ^n$ satisfies
\be \mu ^n  > \left( E_1 \over m_{pl} \right)^n \; , \ee
and the energy dependence in Eq. \ref{bound2} is bounded by
\be
 E_1^{\, \,  n +1 -\beta} \,
\left( 1 - \left(E_2 \over E_1\right)^{1-\beta } \, \right)  \; .
 \label{energ}
\ee
The exponent $n+1-\beta$ is positive for $n > \beta - 1$. The
parameter $\beta$ can vary between $1.6 < \beta < 5$ \cite{Band} (the
lower limit of $\beta \sim 5$ corresponds, however, to bursts which do
not have a high energy tail and are consequently not interesting for
our purposes). For typical bursts $\beta \sim 2.5$.  For $n \geq 2$
and $\beta \sim 2.5$ the exponent $n +1 -\beta$ is positive and
therefore the best bound is obtained by observing in the highest
possible energy band. On the contrary, for $n =1 $ and $\beta \sim
2.5$ the exponent is negative and it is advantageous to use the {\it
  low energy bands}. Note that this analysis is based on the
assumption that the spectrum is given by Eq.  \ref{spectrum}, which is
valid for energies higher than $ {\cal E}_0$. We should therefore
always observe at energies higher than ${\cal E}_0$.

Like the discussion about the optimal redshift, this is another non
intuitive result. Contrary to what would be naively expected, we have
shown that the optimal energy range is not necessarily the highest
one, but depends on the model of symmetry breaking and on the burst
spectrum. These results apply only for a given detector with a fixed
collecting area. Observations in a higher energy band might be
advantageous under different conditions, for instance, if they are
made with a different telescope with a larger area. Finally, cosmic
attenuation sets an upper limit on the energy range. We discuss this
issue in the section \ref{cos-attenuation}.

From Eq. \ref{energ} is evident that the larger the ratio $E_2 / E_1$,
the better the bound (broadening the energy band increases the number
of photons and consequently the telescope resolution); however, in
order to avoid a large spread in the arrival times, the detected
photons in each channel should have comparable energies.

The bound can be rewritten as
\be
\xi_n > \sigma \left[ {{\cal L}_{peak}  \over {\cal L}_*}
\, \, {\cal G}_n (z) \right]^{1/n} \; ,
\label{bound3}
\ee
where
\be \sigma = \left[ {A H_0\over 8 \pi \, b\, m_{pl}^n \, c^2} \,
{\beta - 2 \over \beta -1} \,{\cal L}_* \, {\cal E}_0^{\beta-2} \,
E_1^{n+1-\beta}
 \left( 1 - \left(E_2 \over E_1\right)^{1-\beta} \right)
\right]^{1/n} \; .
\ee
We have introduced the quantity ${\cal L}_* = 6.3 \cdot 10^{51} $
erg s$^{-1}$, which will be useful later on when dealing with
luminosity distributions. In Eq. \ref{bound3} we have explicitly
separated the redshift and burst luminosity and included all the
numerical values and telescope dependent quantities in the
constant $\sigma$,
\be
\sigma \sim 10^{22 {1-n -\beta \over n} }\left(
\,{\beta - 2 \over \beta -1}
\right)^{1/n}
\, \left({A \over 2000 \mbox{ cm}^2}\right)^{1/n} \,
\left( E_1 \over \mbox{MeV}\right) ^{1+{1-\beta \over n}} \,
 \left( 1 - \left(E_2 \over E_1\right)^{1-\beta} \right)^{1/n} .
\ee
We see again that for $n=1$ and $\beta > 2$ going to high energies
does not improve the bound.

The order of magnitude of achievable bounds for typical bursts
obtained in Eq. \ref{bound3} is $10^{-3}$ for $n=1$ and $3 \cdot
10^{-13}$ for $n=2$ (for $\beta = 2.5$ in both cases). This is in
agreement with actual limits found for specific bursts in the
literature. Ellis et al. \cite{Ellis:2002in} used a wavelet
analysis to look for correlations between redshift and spectral
time lags between the arrival times of flares at different
energies, and obtained the bounds of $\xi_1 > 5.6 \cdot 10^{-4} $
and $\xi_2 > 2.4 \cdot 10^{-13}$ at a 95\% of confiance level.
Subsequently this result was improved \cite{Ellis:2005wr} by using
a more complete data set of transient sources with a broad spread
in redshifts to correct for intrinsic time delays (we discuss
intrinsic delays in section \ref{comparacion}). A different
approach was adopted by Boggs et al. \cite{boggs-2004}, who
considered a single and extremely bright burst, GRB 011206. This
burst yielded the bounds of $\xi_1 > 0.015 $ and $\xi_2 > 4.5 \cdot
10^{-12}$ respectively.

\subsection{Cosmic attenuation}

\label{cos-attenuation}

At energies of TeV and higher, the universe becomes opaque due to
the interaction of the gamma ray photons with the background light
to create electron-positron pairs, $\gamma \gamma \rightarrow
\mbox{e}^{-} \mbox{e}^{+}$. The cross section of this reaction is
maximized when the product of the energies of both photons is
$\sim (m_e c^2)^2$. A photon of 10 TeV will interact with an
infrared photon, for instance, creating an electron-positron pair.

\begin{figure}
\begin{center}
\includegraphics{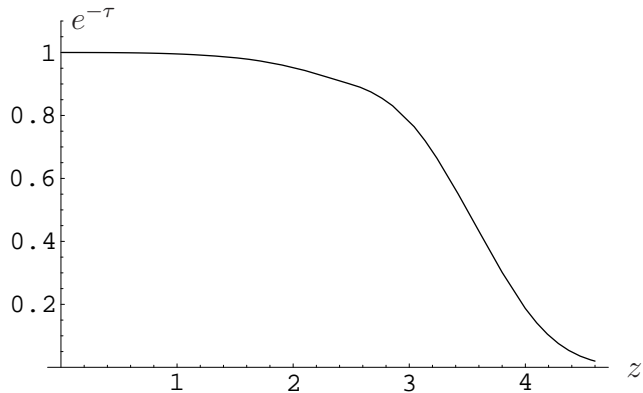}
\put(5,8){$ z$}
\put(-205,140){$ e^{-\tau} $}
\caption{Optical thickness of the intergalactic medium at 10 GeV}
\label{optthick}
\end{center}
\end{figure}

For $n=1$ we are concerned with lower energies, from a few
hundreds keV to a few MeV. A these energies we can safely neglect
attenuation. Fig \ref{optthick} shows the optical thickness of the
extragalactic medium at 10 GeV \cite{Primack:2005rf}. As it can be
seen, attenuation only becomes important at high redshift, $z >3
$. Clearly, it is safe to ignore this effect at lower energies.

\subsection{A quantitative example : GRB 050603A}
\label{quantburst}

For an observed burst on Earth, we can skip Eq. \ref{bound2} which
depends on quantities defined in the rest frame of the burst (${\cal
  L}_{peak}$ and ${\cal E}_0$) and calculate the bound directly from
the flux on Earth. The quantities usually measured are the energy flux
$P(E_1,E_2)$ (in erg cm$^{-2}$ s$^{-1}$) or the photon flux
$P_{\gamma}(E_1,E_2)$ (in photons cm$^{-2}$ s$^{-1}$).

GRB 050603A is a very bright burst with measured redshift $z =
2.821$ observed with two satellites : Konus-Wind (area $\sim 200$
cm$^2$) and {\it Swift}-BAT (area $\sim 5200$ cm$^2$). The
time-integrated spectrum is well fitted by a high energy photon
index $\beta = 2.15 $.
Konus measured a peak flux of $ 3.2 \cdot 10^{-5}$ erg cm$^{-2}$
s$^{-1}$ in the  (20 keV, 3 MeV) band. These values correspond to:
\begin{center}
\begin{tabular}{|c|c|c|}
\hline \multicolumn{2}{|c|}{$\Delta t_{\mbox{{\scriptsize res}}}$
(20 keV,  3 MeV)} &
$\sim  \; 5 \cdot 10^{-4}$ s \\[3pt]
\hline  $n = 1$ &  $\Delta t_{del}$ (20 keV)
& $\sim  \; 9 \cdot 10^{-6}/ \xi \;$  s  \\[3pt]
\hline  $n = 2$ &  $\Delta t_{del}$ (20 keV)
& $\sim  \; 5 \cdot 10^{-29}/ \xi^2 \;$  s  \\[3pt]\hline
\end{tabular}
\end{center}
For the same burst {\it Swift} detected a peak flux of 31.8
photons cm$^{-2}$ s$^{-1}$ in the (15 keV, 350 keV) band. This
implies
\begin{center}
\begin{tabular}{|c|c|c|}\hline
\multicolumn{2}{|c|}{$\Delta t_{\mbox{{\scriptsize res}}}$ (15
keV,  350 keV)} &
$\sim  \; 4 \cdot 10^{-5}$ s \\[3pt]
\hline  $n = 1$ &  $\Delta t_{del}$ (15 keV)
& $\sim  \; 7 \cdot 10^{-6}/ \xi \;$ s  \\[3pt]
\hline  $n = 2$ &  $\Delta t_{del}$ (15 keV)
& $\sim  \; 3 \cdot 10^{-29}/ \xi^2 \;$ s  \\[3pt]\hline
\end{tabular}
\end{center}

\noindent If the time delays at these low energy bands were not
contaminated with intrinsic spectral delays at the source, this
burst could have led to the following limits:

\begin{center}
\begin{tabular}{|c|c|c|} \hline
&  Konus  &{\it Swift}\\[3pt] \hline
$n = 1$ & $\xi > 0.02$ &$\xi > 0.2 $ \\[3pt] \hline
$n = 2$ & $ \xi > 3 \cdot 10^{-13}$  &$ \xi > 9 \cdot 10^{-13}$ \\[3pt] \hline
\end{tabular}
\end{center}

The time resolution used in this example, $4 \cdot 10^{-5}$ s, is in
fact below the limiting time resolution of {\it Swift}, 0.1 ms.  {\it
  Swift} resolution would have led to a slightly weaker bound, $\xi_1
> 0.07 $. Furthermore, we have considered here photons all the way
down to 15 keV ({\it Swift}) or 20 keV (Konus). Limiting the
discussion to photons above $\sim 100$ keV, for which Eq.
\ref{spectrum} is accurate, would have reduced the limits further.

The numbers obtained here are clearly idealized and should serve only
as an example of what can be achieved. In order to set an experimental
bound, we need to observe the burst in two different energy channels
and compare the arrival time in each one. An additional problem is
posed by the observed intrinsic delay in the emission times of the
photons at the source \cite{Ellis:2002in,Piran:2004qe} which we
discuss in section \ref{comparacion}.

\section{Distribution of bursts}

GRB 050603A could have given a very powerful bound on $\xi$. But how
likely would it be to detect a burst yielding such a bound or a higher
one ? From the results of the previous sections, the more luminous
and closer the burst, the stronger the bound. In this section we
estimate the probability of finding such a burst, given an empirical
luminosity and space distribution of bursts.

\begin{figure}
\begin{center}
\includegraphics{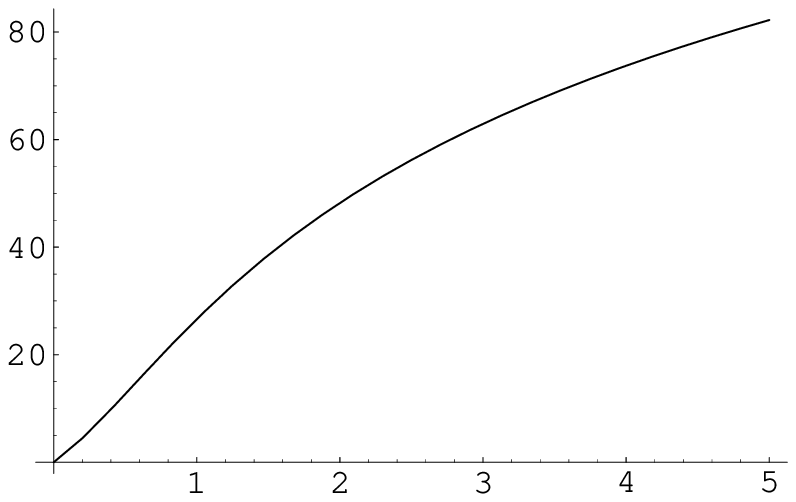}
\put(5,8){$ z$}
\put(-220,160){${\cal L} / {\cal L}_* $}
\put(-170,120){$\xi > \xi_0$}
\put(-90,60){$\xi < \xi_0 $}
\caption{The luminosity and the redshift needed to give a bound
of $\xi_0 =0.2 $ for $n=1$. Higher bounds are obtained above the curve.}
\label{contplot}
\end{center}
\end{figure}

To improve a bound, $\xi_0$, we need to detect a burst with a
luminosity and a redshift such that
\be
\sigma^n \, \, {{\cal L}  \over {\cal L}_*}  \,
\, {\cal G}_n (z) > \xi_0^n  \; \; .
\label{cond-improving}
\ee
Eq. \ref{cond-improving} define a region in the luminosity and
redshift space-phase (see Fig. \ref{contplot}). The probability for
such a burst to happen is given by the integral of the probability of
finding a burst over this region.

Let us introduce the local peak luminosity function, $\Phi_0(\L)$,
defined as the fraction of GRBs with luminosities in the interval
$\log \L$ and $\log \L + d \log \L$, can be approximated by
\cite{schmidt-99}:
\begin{equation}
\label{Lfun}
\Phi_0(\L)=c_0
\left\{ \begin{array}{ll}
(\L/\L_*)^{{\alpha_1}} &  \L_*/\Delta_1 < \L < \L_* \\
(\L/\L_*)^{{\alpha_2}} & \L_* < \L < \Delta_2 \L_*
\end{array}
\right. \;,
\end{equation}
where $c_0$ is a normalization constant such that the integral over
the luminosity function equals the unity.

There is a strong evidence that long GRBs follow the comoving star
formation rate (SFR), $ R_{SFR}(z)$.  Namely, $R_{GRB}(z)$, the
comoving GRB rate satisfies $R_{GRB}(z) \propto R_{SFR}(z) $. We
employ the parametrization of Porciani et al. \cite{porciani-01} for
the comoving SFR distribution. From it we write
\begin{eqnarray}
\label{SFR}
R_{GRB}(z) =  \rho_0 \frac{23
\exp(3.4z)}{\exp(3.4z)+22}
\,  { \sqrt{ \Omega_m  (1+z)^3 +  \Omega_\Lambda}\over (1+z)^{3/2}} \; .
\end{eqnarray}
The luminosity function at redshift $z$ is therefore $\Phi(z,\L) =
\Phi_0(\L) \,R_{GRB}(z)$.  Guetta et al. \cite{Guetta:2003zp} used
the BATSE peak flux distribution, to estimate the parameters
$\rho_0$, $\alpha_1$ and $\alpha_2$. For long bursts they found two
different fits:

\begin{center}
\begin{tabular}{ccccccc}
\hline & $\alpha_1$ & $\alpha_2$ & $\Delta_1$ & $\Delta_2$
&${\cal L}_*$ (erg s$^{-1}$) & $ \rho_0$ (Gpc$^{-3}$ yr$^{-1}$)\\
\hline
I & -0.1 & -2 & 30 & 50 & $6.3 \cdot 10^{51}$ & 0.18 \\
II & -0.6 & -3 & 30 & 50 & $1.6 \cdot 10^{52}$& 0.16 \\
\hline
\end{tabular}
\end{center}
Short GRBs, which constitute about one quarter of the observed
bursts, do not follow the SFR \cite{GP05, GP06} and will not be
discussed here.

The probability of detecting a burst which sets a bound $\xi > \xi_0$
is
\be
N(\xi > \xi_0 ) = \int ^\infty_0 {R_{GRB}(z) \over 1+z } {d V(z) \over dz} dz
\int^\infty_\Lambda
 \Phi_0(\L) d\log \L ,
\label{prob-num}
\ee
where the factor $(1+z)^{-1}$ accounts for the cosmological time dilation and
$V(z) $ is the comoving volume. The factor $\Lambda$ is defined as
\be
\Lambda = \max \left( {\xi_0^n \, {\cal L}_* \over \sigma^n {\cal G}_n (z)} ,
{\cal L}_{min} \right) ,
\ee
where ${\cal L}_{min}$ is the minimal luminosity for a burst to be
detected
\be
{\cal L}_{min} =  4 \pi \, d(z)^2 \, (1+z)^\beta \, {\beta - 1
  \over \beta -2} \, { {\cal E}_0^{2-\beta} \over E_1^{1-\beta} - E_2^{1-\beta} } \,{\cal
N}_{min} . \ee
${\cal L}_{min}$ depends on the sensitivity of the telescope, ${\cal
N}_{min}$, which is the minimal photon flux necessary to trigger the
instrument. We considered for $n=1$ an idealized detector, Det. I, with an area
of 5200 cm$^{2}$ making observations in the energy band (500 keV,
2 MeV). For $n=2$, we focused on the forthcoming spatial observatory
GLAST which will be launched in 2006. We considered the energy band
(100 MeV, 1 GeV) where GLAST is expected to have an effective area of
8000 cm$^{2}$.

Integrating numerically Eq. \ref{prob-num} we find

\begin{center}
\hspace*{-3cm}
\begin{tabular}{c|l|c|r|c|}
 \cline{2- 5}
& \multicolumn{2}{c|}{ {\bf Det. I}} & \multicolumn{2}{c|}{ {\bf GLAST}} \\
\cline{2- 5}
& $\;\; \; \; \xi_1 $ & rate (bursts yr$^{-1}$) & $\xi_2 \;\; \;\;\; \; \;$
& rate (bursts yr$^{-1}$) \\ \cline{2- 5}
{\bf GRB 021206} & {\bf 0.015} &(7.4 - 6.9)$\cdot 10^{-1}$ &
${\bf 4.5 \cdot 10^{-12}}$  & 9.4 - 12.8 \\  \cline{2- 5}
& 0.05 & (2.8 - 3.1)$\cdot 10^{-1}$ & $7 \cdot 10^{-12}$
& 1.9 - 2.2 \\ \cline{2- 5}
& 0.1 & (9.1 - 6.9)$\cdot 10^{-2}$ & $ 10^{-11}$
&  (4.8 - 3.5)$\cdot 10^{-1}$ \\ \cline{2- 5}
& 0.5 & (2.1 - 1.3)$\cdot 10^{-3}$ & $2.5 \cdot 10^{-11}$ &
 (5.0 - 2.4)$\cdot 10^{-3}$ \\ \cline{2- 5}
& 1.0& (3.5 - 2.0)$\cdot 10^{-4}$ & $ 5\cdot 10^{-11}$
& (7.2 - 3.9)$\cdot 10^{-5}$\\ \cline{2- 5}
\end{tabular}
\end{center}

\noindent The values in the first row correspond to the bounds
obtained from GRB 021206 \cite{boggs-2004}. To estimate ${\cal
  L}_{min}$, we assumed that the ideal detector, Det. I, has a
sensitivity of ${\cal N}_{min} \sim $ 1 ph cm$^{-2}$ s$^{-1}$, which
is comparable to the sensitivities estimated by Band
\cite{Band:2002te} for several detectors in similar energy bands. For
GLAST, we took ${\cal N}_{min} \sim 4 \cdot 10^{-6}$ ph cm$^{-2}$
s$^{-1}$, which roughly means that the detector is very quiet at these
energies and a detection of 6 photons during the whole duration of the
burst ($\sim$ 3 minutes) is enough to identify it.

Finally we took into account the partial sky coverage of any real
telescope to compute the number of bursts observed per year.  GLAST
opening angle will be $\sim $ 2 stereoradian; we used a similar
opening angle for our idealized detector.

\section{Comparison with other works}

\label{comparacion}

The idea of using GRBs to set experimental bounds on a possible
violation of Lorentz symmetry  was first suggested by
Amelino-Camelia et al. \cite{Amelino-Camelia:1997gz}. Later on,
several groups made use of it to explore these limits
\cite{Ellis:2002in,boggs-2004,Ellis:2005wr}. The current best
bounds have been obtained by Boggs et al. \cite{boggs-2004} who
used a very bright burst, GRB 021206, to set a limit on the
symmetry breaking scale. The data consisted of light curves in six
energy bands spanning 0.2 - 17 MeV.  The redshift of this burst is
not known, but an approximated redshift of $z \simeq 0.3$ was
estimated from the spectral and temporal properties of the burst
(this method involves, however, a very high uncertainty which can
be as high as a factor of 2).

The observed fluence of GRB 021206 is $1.6 \cdot 10^{-4}$ ergs
cm$^2$ at the energy range of 25-100 kev \cite{Hurley02}. This
puts GRB 021206 as one of the most powerful bursts ever observed.
GRB 021206 also shows a very atypical photon spectrum at the MeV
range. Instead of decreasing with the energy following a power law
with $\beta \sim 2.5$, it is almost flat from 1 MeV up to 17 MeV,
namely $\beta \sim 0$ (this implies in particular that $F_{\nu}$
increases with $\nu$ in these energies). This flatness allows to
resolve a fast flare and to determine its peak time and
uncertainty in several bands. The analysis of the dispersion of
these peak times yields to the lower bounds $\xi_1 > 0.015$
and $\xi_2 > 5 \cdot 10^{-12}$ \cite{boggs-2004}. Applying our
method on the energy band 15-350 keV using data from the GRB
050603A, we obtained a theoretical upper limit to the lower bounds
of $\xi_1 > 0.2 $ and $\xi_2 > 9 \cdot 10^{-13}$. These numbers
represent the best bounds that  could be obtained if the time
resolution of the detector was high enough ($ \sim 5 \cdot
10^{-5}$ s), the detector noise was negligible and we had at our
disposal the light curves in at least two energy channels.

Our conclusions on the optimal redshift and energy band are based on a
power law spectrum $E^{-\beta}$ with $\beta \geq 2$.  They arise from
comparing the time delay, which always increases with the energy, with
the time resolution of the telescope.  Comparing both energy
dependencies, we found in section \ref{conditions-energy-redshift}
that for $n=1$ and $\beta >2$ is better to observe at low energies.
However if $\beta < 2$, like in GRB 021206 in the MeV range, this
conclusion does not hold and it is preferable to use the highest
available energy band. In this case the GLAST Large Area Telescope
will be a very powerful tool. It is expected to be sensitive from 20
MeV to 300 GeV with a peak effective area in the range 1-10 GeV of
8000 cm$^2$. Observing with GLAST in energy bands below 10 GeV where
cosmic extinction is still negligible (see fig.  \ref{optthick}) can
improve dramatically our current bounds.  At present little is known
about GRB emission at energies higher than 50 MeV and therefore it is
not possible to estimate how common are bursts with $\beta < 2$.
Taking advantage of atypical bursts to explore even higher energies is
an exciting possibility to keep in mind, however at present it is
difficult to design a strategy based only in these bursts.

As already mentioned, our bound must be interpreted as a
theoretical one, {\it i.e.} the highest bound that could be set,
were the best conditions achieved. We already commented on the
necessity of observing in at least two energy channels and to take
into account the real time resolution of the detector.

An additional serious problem is the intrinsic lack of
simultaneity in the pulse emission in the keV regime
\cite{Ellis:2002in,Piran:2004qe}. Soft emission has a time delay
relative to high energy emission \cite{Norris:96}. While the
reason for this phenomenon is not understood, an anti-correlation
between the spectral evolution timescale and the peak luminosity
has been found \cite{Norris:1999ni}. There are two different ways
to deal with the intrinsic delay. The first is to try to reduce it
by choosing very luminous bursts and observing in MeV or higher,
where the delay, if still exists, seems to be smaller. The second
approach is based on the fact that the delays produced by a
violation of Lorentz symmetry increase with the redshift of the
source, whereas intrinsic time delays are independent of the
redshift of the source \cite{Ellis:2005wr}. Thus, a systematic
comparison of a delays in a group of bursts with known redshifts
could enable us to distinguish between intrinsic and redshift
dependent delays.  Using a sample of 35 bursts with known
redshifts, Ellis et al. \cite{Ellis:2005wr} established a lower
limit of $\xi_1 > 7 \cdot 10^{-4}$ on the symmetry breaking scale.
These bounds are two orders of magnitude lower than our
theoretical limits. This difference demonstrates the importance
that intrinsic time delays, noise and the real instrumental
resolution can have.

\section{Conclusions}

\label{conclusion}

Our goal was to explore the potential of GRBs to set bounds on
Lorentz violation and to find optimal techniques to do so.  We
modified the dispersion relation of photons by adding an extra
term proportional to the photon momentum to the power $n+2$. We
have shown that in models with $n=1$ it is possible to explore
energies which are close to the Planck energy. When $n=2$, the
energies explored are smaller, around $10^7 $ GeV.  These bounds
are idealized and they do not take into account experimental
limitations or the intrinsic time-structure of the $\gamma$-ray
emission. They should serve as theoretical estimations of what can
be achieved. The methodology we use here can be used to design
future optimal experiments for observing this effect (or setting
bounds on it).

We have modelled the burst high energy emission with a power law
spectrum $E^{-\beta}$ with $\beta \geq 2$. This fits well the time
integrated emission of most of the bursts. We found two non intuitive
results: (i) The optimal redshift to set the strongest bound is less
than 1.  (ii) For $n=1$, low energy, rather than high energy emission
is preferred. Both results are counter-intuitive since the Lorentz
violation delay increases with the distance and with the energy.
However, distance or observations at high energies (where the flux is
lower) dilute the photons reducing the temporal resolution achieved on
Earth. It turns out that this is the dominant effect.

In the models with $n=2$, going to higher energies always improves
the bounds. Here the situation will be remarkably changed when the
spatial observatory GLAST will become operational.

We have also investigated the probability of improving the current
experimental bounds, given a phenomenological luminosity and space
distribution of bursts.  As we are discussing idealized bounds,
this probability should only be trusted up to an order of
magnitude.

\section*{Acknowledgments}

We would like to thank Steven Boggs, David Palmer, Joel Primack,
David Smith and Raquel de los Reyes L\'opez. We especially thank
Matthew Kleban for many useful discussions and a critical reading
of this manuscript. This research was supported by the EU-RTN
``GRBs - Enigma and a Tool'' and by the Schwarzmann university
chair (TP).

\end{document}